\documentclass[pra,twocolumn,aps]{revtex4}
\usepackage{amsfonts}
\usepackage{amsmath}
\usepackage{graphicx}
\usepackage{sidecap}

\begin{document}

\title{Single-Photon Superradiance in Cold Atoms}

\author{Rafael A. de Oliveira$^{1}$, Milrian S. Mendes$^1$, Weliton S. Martins$^1$, Pablo L. Saldanha$^{2}$, Jos{\'e} W. R. Tabosa$^{1}$, and Daniel Felinto$^{1}$}

\affiliation{$^{1}$Departamento de F\'{\i}sica, Universidade Federal de Pernambuco, 50670-901 Recife-PE, Brazil \\ $^{2}$Departamento de F\'isica, Universidade Federal de Minas Gerais, 30161-970 Belo Horizonte-MG, Brazil }

\begin{abstract}
The interaction of an ensemble of atoms with common vacuum modes may lead to an enhanced emission into these modes. This phenomenon, known as superradiance, highlights the coherent nature of spontaneous emission, resulting in macroscopic entangled states in mundane situations. The complexity of the typical observations of superradiance, however, masks its quantum nature, allowing alternative classical interpretations. Here we stress how this picture changed with the implementation ten years ago of a new process for single-photon generation from atomic ensembles. We present then the last piece of evidence for the superradiant nature of such process, reporting the observation of an accelerated emission of the photon with a rate that may be tuned by controllably changing the number of atoms in the ensemble. Such investigation paves the way to a new, bottom-up approach to the study of superradiance.
\end{abstract}

\maketitle

\section{Introduction}

Sixty years ago, it was pointed out by Robert Dicke that the full quantum mechanical treatment of spontaneous emission from an ensemble of atoms could lead to enhanced, ``superradiant'' emissions in particular modes~\cite{Dicke1954}. Such effect would result in the medium spontaneously radiating in a burst much more directional and faster than if the atoms were emitting independently. This strong cooperativity is originated in the coherent nature of the interaction between atoms and vacuum and in the fact that some vacuum modes are coupled to the whole ensemble. Over time, superradiance has attracted great attention, since it may occur at common situations in many different systems and is the manifestation of a macroscopic entangled state spontaneously generated in the medium.

Even though, superradiance is still an effect hard to characterize and isolate. The interaction between particles and the induction of macroscopic polarizations in the sample, for example, may destroy or mask the observations of spontaneous cooperativity~\cite{Gross1982}. The first observations of strong superradiance, reported in the 1970s using extended ensembles~\cite{Skribanowitz1973,Gross1976}, were mixed with strong propagation and diffraction effects~\cite{Gross1982,MacGillivray1976}. On top of that, there is the question of the necessity or not of the full quantum mechanical treatment, and the entangled states that naturally comes with it, to understand the experimental results. It is clear that various aspects of superradiance have classical analogues~\cite{Dicke1954,Gross1982,Skribanowitz1973}, like the enhanced decay in a burst, which can be obtained from an ensemble of antennas emitting in phase.

However, a new process for generating photon pairs from atomic ensembles, proposed in 2001~\cite{Duan2001} and implemented two years later~\cite{Kuzmich2003}, significantly moves away from the semiclassical views of superradiance and reinforces the central role of macroscopic entanglement for the understanding of the phenomenon and of its potential applications. This process was part of a broad protocol for quantum communication over long distances, known as the Duan-Lukin-Cirac-Zoller (DLCZ) protocol. In the following, we highlight the impact the experimental implementation of the DLCZ protocol had on the study of superradiance, and report an investigation on the dynamics of such photon-pair generation that directly reveals its superradiant character. Our measurements of acceleration in the radiation process complement the known collective enhancement in directionality in the system. In this way, all key aspects of the phenomenology of superradiance have now been identified in connection to the DLCZ photon-pair-generation process, which can be employed henceforth as a framework for the systematic study of superradiance itself. 

Below, in Sec.~\ref{II} we discuss the role of superradiance in the DLCZ protocol, as we introduce the basic process behind single-photon generation in our system. Section~\ref{III} presents our experimental setup, its characterization, and our first results related to superradiance. In this section, we measure the threshold optical depth of the atomic ensemble for superradiant behavior, as revealed by a sharp growth of the probability to detect the emitted single photon in the correct mode, combined with the appearance of nonclassical correlations in the system. We also demonstrate that such detection probability grows, at threshold, with the square of the number of atoms in the ensemble. Section~\ref{IV} focus on our observations of Rabi oscillations for the collective atomic mode and their comparison to an analytical theory for the reading process. This comparison is the basis to extract the decay time of the excited state modified by the condition of superradiance. In Sec.~\ref{V} we demonstrate experimentally, then, that such decay time is decreased as the number of atoms in the ensemble grows in the way expected for a superradiance process. Section~\ref{VI} provides finally our conclusions and perspectives on the subject.

\newpage

\section{Superradiance in the DLCZ protocol}
\label{II}

\begin{figure}[htb]
  \hspace{-0.5cm}\includegraphics[width=9.2cm]{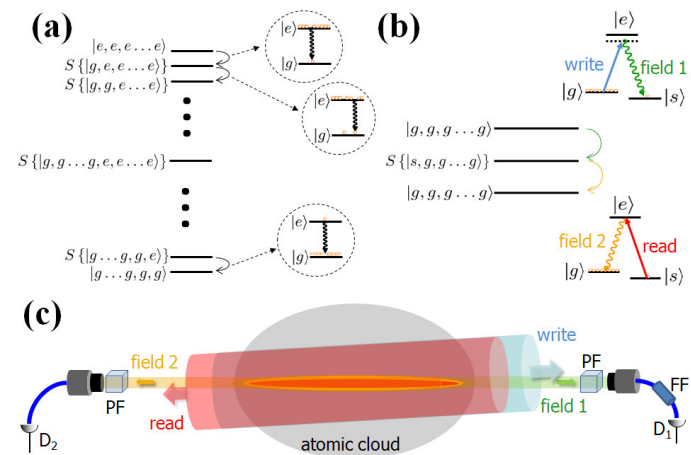}
  \vspace{-1.2cm}
  \caption{(a) Typical superradiant cascade in an ensemble of atoms with two levels, $|g\rangle$ and $|e\rangle$. $S\{ | \cdots \rangle \}$ indicates symmetrical entangled states generated along the cascade. (b) Minimal superradiant cascade of the DLCZ protocol, involving atoms with three levels: $|g\rangle$, $|s\rangle$, $|e\rangle$. Spontaneous emissions on fields 1,2 are observed after excitation of the ensemble by write,read pulses, respectively. (c) Spatial configuration for fields in (b). Write and read are large beams counterpropagating to each other. Fields 1,2 are defined by the optical fibers (in blue) carrying them to detectors $D_1$,$D_2$. The detected modes, forming a small angle with the excitation fields, define the region in the atomic cloud that stores the collective entangled state. PF and FF denote polarization and frequency filters, respectively.}\label{fig1}
\end{figure}

In order to understand its classical analogy, one may view superradiance as a cascade of emissions starting at a state of maximum energy of an ensemble of two-level atoms~\cite{Dicke1954,Gross1982}, see Fig.~\ref{fig1}a. Labeling $|g\rangle$ and $|e\rangle$ the ground and excited states, respectively, the initial state $|e,e,e \cdots e\rangle$ of the ensemble would carry then no coherence between the atoms. Once a first atom spontaneously decays and emits a photon in the common mode, the system is left in a large symmetrical collective state $S\left\{ |g,e,e \cdots e\rangle \right\} \propto \sum_i |e,e \cdots g_i \cdots e, e\rangle$, for which an atom $i$ decayed, but it is not known which one. This large entangled state has a fixed phase between its parts and some coherence, coming from the small probability each atom has of being in the ground state. As the cascade proceeds down the energy ladder, new symmetrical states are formed with more and more atoms in $|g\rangle$. The amount of coherence then grows in the sample. After many emissions, the system's behavior is dominated by its large coherence and hereafter may be approximated by its classical analogue. In the end, the ensemble is left in state $|g,g,g\cdots g\rangle$, with again no coherence. In the picture presented above, which have been quite successful for explaining experimental observations~\cite{Gross1982,Skribanowitz1973}, the full quantum mechanical treatment is only required to describe the spontaneous trigger for an otherwise classical decay process. 

\vspace*{-0.6cm}

As for the DLCZ protocol, its building block is the generation of macroscopic entangled states that can be stored over long times. In order to do so, three-level atoms in $\Lambda$ configuration are employed (Fig.~\ref{fig1}b), with an extra ground state $|s\rangle$ added to the above picture. All atoms are initially prepared in $|g\rangle$. The sample is then excited by a laser pulse ({\it write}) detuned from the transition $|g\rangle \rightarrow |e\rangle$ and, with small probability, a photon may be spontaneously emitted in the transition $|e\rangle \rightarrow |s\rangle$, with an atom being simultaneously transferred to $|s\rangle$. If the photon is emitted in a mode ({\it field 1}) common to the ensemble, its detection heralds the preparation of the symmetrical entangled state $S\{ |s,g,g\cdots g\rangle\}$~\cite{Duan2001}. A second laser pulse ({\it read}) may now excite the transition $|s\rangle \rightarrow |e\rangle$ and, with high probability, a second photon ({\it field 2}) is emitted in the transition $|e\rangle \rightarrow |g\rangle$. In the end, all atoms are left again in $|g\rangle$. This last part of the process is the one connected to the usual phenomenology of superradiance, the high probability for the second emission coming from its strong directionality~\cite{Duan2001}. Differently from previous experiments in superradiance, however, a strong read pulse may ``open'' the medium to the outgoing photon, using the effect known as Electromagnetically Induced Transparency (EIT)~\cite{Boller1991}, reducing considerably distortions due to propagation of the photon through a thick, extended sample.

After the first implementations of this process~\cite{Kuzmich2003,Chou2004,Eisaman2004}, a major development was the introduction in 2005 of a Four-Wave-Mixing configuration (Fig.~\ref{fig1}c) for the photon pair generation~\cite{Balic2005,Matsukevich2005}, which solved most complications related to diffraction in the superradiant emission. The write pulse have here a considerably larger waist for its transversal mode than the one for field 1, which is fiber-coupled and detected with a small angle to the direction of the write beam. In this case, the stored state with a single excitation in the ensemble would be given by $|1_{at} \rangle = \sum_{i} A_i |g,g \cdots s_i \cdots g\rangle$, where $A_i$ gives the probability that the $i$th atom contributed to the detected mode~\cite{Mendes2013}. In this way, the optical fiber for field 1 defines the spatial shape of the collective state stored in the ensemble. If the read field has also a large waist and is counterpropagating to the write beam, then field 2 is generated in the conjugated mode to field 1~\cite{Mendes2013}. The result is a superradiant emission of photon 2 in a well defined single mode, which can be coupled to an optical fiber with high efficiency~\cite{Laurat2006}. An indirect observation of the superradiant increase of the spontaneous decay rate of the system was also reported in~\cite{Mendes2013}, as part of a detailed study of the saturation and spectrum of the readout process of field 2 under conditions of strong decoherence due to inhomogeneous magnetic fields acting on the atomic ensemble.

The overall process described above amounts then to a minimal superradiant cascade (Fig.~\ref{fig1}b), in which a single photon is responsible for the preparation of the initial macroscopic entangled state that later results in the superradiant emission of another single photon. The fundamental importance of a minimal, single-photon superradiance have been emphasized in recent years in a number of publications~\cite{Scully2006,Scully2009,Bienaime2013}, and other experiments at the single-excitation level have been reported in studies of nuclear scattering of synchrotron radiation~\cite{Rohlsberger2010} and of two-photon cascade transitions in cold atoms~\cite{Chaneliere2006}. The essential quantum-mechanical nature of the effect, in this case, can be directly apprehended from the interplay between its wave-like (collective interference) and particle-like (single-photon detection) aspects. The DLCZ protocol, however, adds to this picture the possibility to generate complex entangled states between different atomic ensembles~\cite{Chou2005,Chou2007,Chen2010,Choi2010} and even to explore these states for practical applications~\cite{Duan2001}.  

\section{nonlinear enhancement and threshold}
\label{III}

\begin{figure}[htb]
  \hspace{0.0cm}\includegraphics[width=8.5cm]{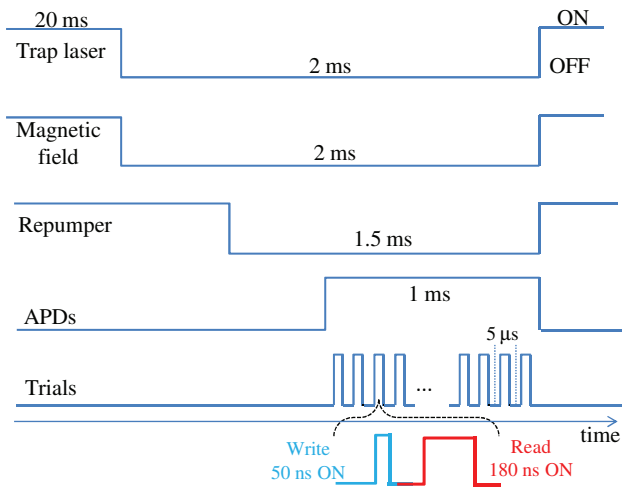}
  \vspace{-0.5cm}
  \caption{Timing for first series of measurements on the threshold for superradiance.} \label{figS1}
\end{figure}

In our experimental setup, the atomic ensemble is a cloud of cold cesium atoms obtained from a magneto-optical trap, with the trap laser tuned 15~MHz below the $6S_{1/2}(F=4) \rightarrow 6P_{3/2}(F^{\prime}=5$) transition and the repumper laser resonant with the $6S_{1/2}(F=3) \rightarrow 6P_{3/2}(F^{\prime}=4)$ transition. Levels $|g\rangle$, $|s\rangle$ and $|e\rangle$ are given by the hyperfine states $|6S_{1/2}(F=4)\rangle$, $|6S_{1/2}(F=3)\rangle$ and $|6P_{3/2}(F^{\prime} = 4)\rangle$, respectively. The trap laser is kept on for 20~ms and, together with the trap's quadrupolar magnetic field, turned off for 2~ms, see Fig.~\ref{figS1}. During this 2~ms period, the repumper laser is kept on for an extra 0.5~ms, in order to help preparing the atomic ensemble with all atoms initially at $|g\rangle$. The avalanche photodetectors (APDs) for the photons are then turned on for 1~ms in the last portion of the 2~ms interval. They have $45\%$ detection efficiency for photons around 850~nm.

\vspace{-0.0cm}

During the 1~ms the APDs are on, a sequence of write and read pulses with durations of 50 and 180~ns, respectively, is sent every 5~$\mu$s to excite the ensemble (Fig.~\ref{figS1}). The exciting pulses are cut from cw diode lasers using acousto-optic modulators with $\approx\,$20~ns rise time. The read pulse arrives in the ensemble 50~ns after the end of the write pulse. The write field was weak and detuned 35~MHz below resonance, to avoid absorption in order to guarantee an uniform excitation of the ensemble. The read pulse was strong and tuned to resonance. In this way, it performed a double role, completely reading out the stored excitation and optically pumping the atoms back to $|g\rangle$. We employ the geometrical configuration of Fig.~\ref{fig1}c, with 400~$\mu$m and 200~$\mu$m for the diameters of the transversal modes of write/read and field 1/field 2, respectively, and a 2$^{\circ}$ angle between them. The ensemble is about 3~mm long. Write and read fields have linear polarizations opposite to each other and to the respective photons they generate. Polarization filters were placed then in front of the fibers for fields 1,2 (Fig.~\ref{fig1}c) to separate the photons from their respective excitation pulses. For field 1, a frequency filter is employed to eliminate photons from spontaneous decays in the transition $|e\rangle \rightarrow |g\rangle$. This filter consists of an in-fiber Fabry-Perot with free spectral range of 20~GHz and linewidth of 400~MHz (FWHM). 

\begin{figure}[htb]
  \hspace{0.0cm}\includegraphics[width=8.0cm]{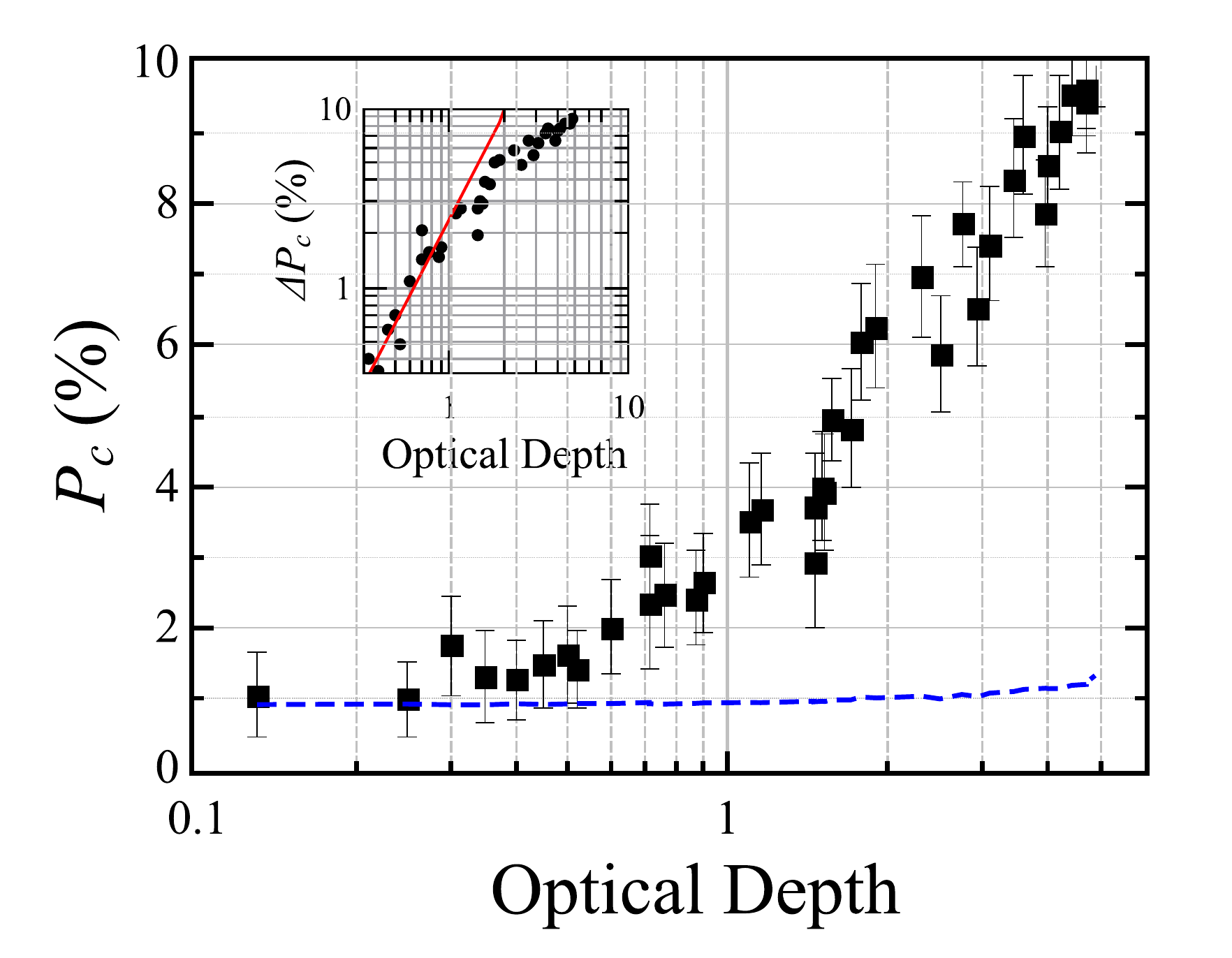}
  \vspace{-0.5cm}
  \caption{Squares provide the total conditional probability $P_c$  for extracting the second photon, once the first was detected, as a function of the ensemble's optical depth. The dashed blue line connects the respective results for $P_2$, the unconditional probability to detect a photon in field 2. The employed read pulse was 180~ns long. The inset provides a log-log graph for $\Delta P_c = P_c - P_2$ versus $OD$. The red line is a liner fit of slope $s = 1.9\pm 0.3$, such that $\Delta P_c \propto OD^s$.}\label{fig2}
\end{figure}

In order to characterize the single-photon regime of field 2, both detectors in Fig.~\ref{fig1}c were substituted by pairs of detectors connected to the output of fiber beamsplitters~\cite{Mendes2013}. In this way, we are able to measure the integrated quantities $P_i$ and $P_{ij}$ giving, respectively, the probability of having single detections in field $i$ and the probability of having joint detections in fields $i,j$. The total probability of having a detection in field 2 conditioned to one in field 1 is then given by $P_c = P_{12}/P_1$. A direct observation of the onset of collective enhancement in the system as the number of atoms $N$ increases is plotted in Fig.~\ref{fig2}, through the dependence of $P_c$ (squares) with the sample's Optical Depth $OD$ in the transition $|g\rangle \rightarrow |e\rangle$, which is proportional to $N$. 

$OD$ was determined as in Ref.~\cite{Mendes2013}, and changed by tuning the power of the trap laser.  Our standard way to measure the Optical Depth of the atomic ensemble, then, was to send through it a weak, long pulse (about 0.5-$\mu$s long) resonant to the $|g\rangle \rightarrow |e\rangle$ transition. Comparing the signal for the center of the pulse after the cell with ($V_f$) and without ($V_i$) the atomic cloud on the pathway, we could calculate directly the optical depth of the cloud through the expression $OD = - \ln (V_f/V_i)$. We obtained the same results for $OD$ if we tuned the pulse over the resonance and fitted the results with a Lorentzian profile. The measurement of $OD$ was performed typically in the center of the interval the APDs were on (see Fig.~\ref{figS1}), without write or read fields acting on the ensemble. 

The single-photon character of field 2 was demonstrated by measuring a significant decrease of the quantity $P_{cc} = P_{122}/P_1$ with respect to what is expected for coherent fields. $P_{122}$ is the probability for a triple joint detection with two detections in field 2 following one in field 1. In this way, $P_{cc}$ is the conditional probability of having two detections in field 2 after one in field 1. Operationally, the single-photon character results in $g_2^c = P_{cc}/P_c^2 < 1$, with $g_2^{c}$ the second order auto-correlation function for the conditioned field 2. For the largest $OD$s in Fig.~\ref{fig2}, we obtain $g_2^c = 0.23 \pm 0.06$. The single-photon character may be indicated also by an indirect measurement, through the quantity $P_c/P_2$, with $P_2$ the unconditional probability to detect a photon in field 2~\cite{Laurat2006}. This quantity measures directly the correlation between the photons, since it quantifies how much a detection in field 1 increases the probability of obtaining another in field 2 in the same trial. For our system, $P_c > 2P_2$ indicates the presence of quantum correlations~\cite{Kuzmich2003}, and $P_c = P_2$ indicates no correlations at all.

\vspace*{-0.0cm}

\begin{figure}[htb]
  \hspace{0.0cm}\includegraphics[width=8.0cm]{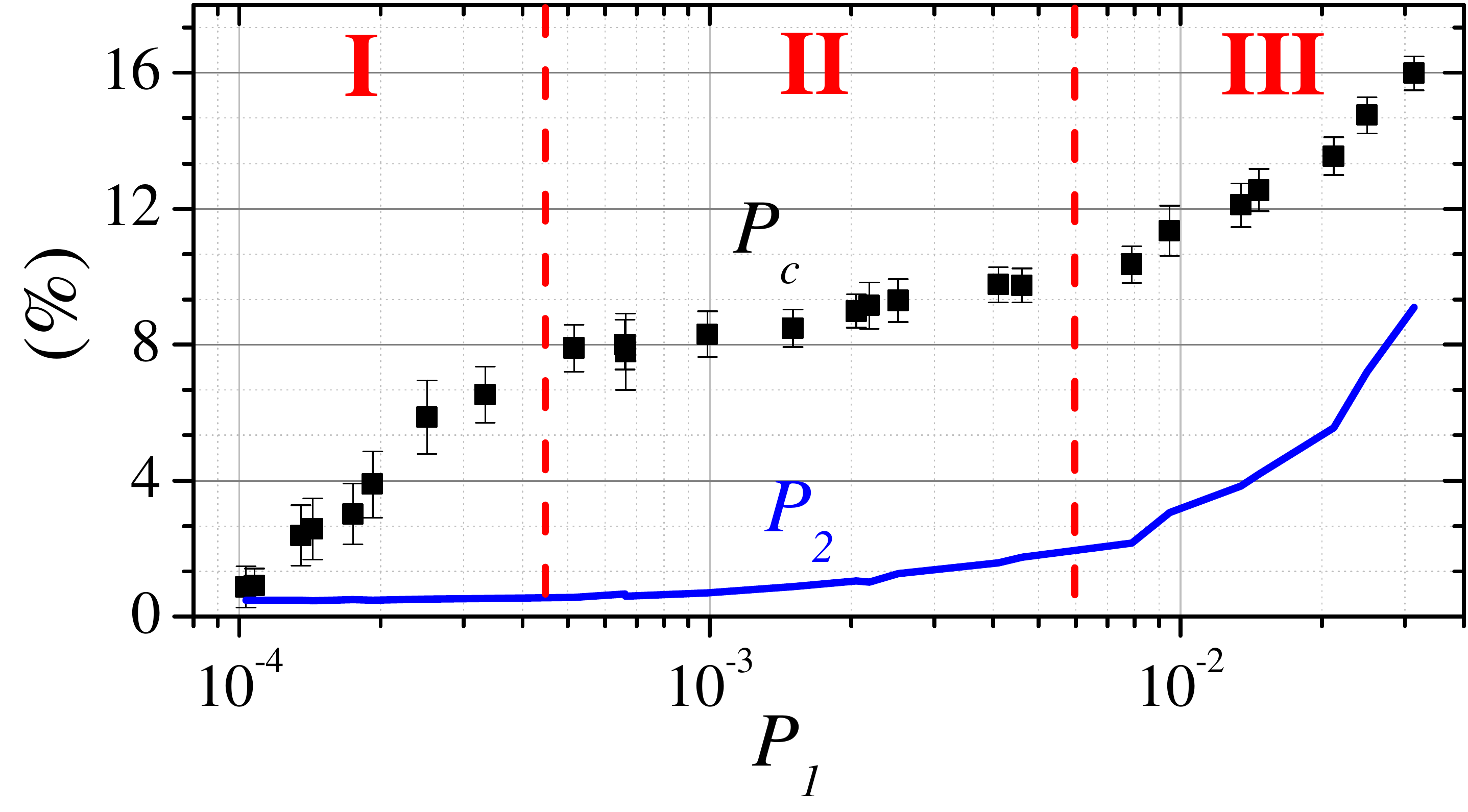}
  \vspace{-0.4cm}
  \caption{Conditional $P_c$ and unconditional $P_2$ probability of detecting the second photon as a function of the probability $P_1$ of detecting the first photon.} \label{figS4}
\end{figure}

There are other straightforward ways to verify the single-photon character of field 2. For the DLCZ photon-pair-generation process, the conditional preparation of a single stored excitation would lead to a $P_c$ value independent of $P_1$~\cite{Laurat2006}. In Fig.~\ref{figS4}  we plot then $P_c$ versus $P_1$ for a variation of more than two orders of magnitude in $P_1$. We can clearly distinguish region II as the single-photon regime~\cite{Laurat2006}, where all the data in the manuscript was taken. Region I is dominated by the noise in field 1, which decreases $P_c$ due to spurious detections in field 1. Region III is the multi-photon regime, where more than one excitation are stored in the ensemble as a result of a strong write pulse.

The plateau on region II indicates $\approx 9\%$ probability to detect a field-2 single photon. In order to obtain an estimation for the probability to extract the single photon from the ensemble in the correct mode, one has to estimate all losses in the detection channel from the face of the ensemble up to the detectors. In our case, we have an 8\% loss associated to the non-coated windows of our vacuum chamber, a 45\% detection efficiency, and about 55\% of other losses in the transmission from the output of the vacuum chamber up to the face of the detectors. After all, we have a net efficiency of about 19\% for transforming a single photon in the output of the ensemble into a click of our field-2 detector. The observed plateau at 9\% indicates then a probability of $\approx 47\%$ of extracting the single photon from the ensemble in the correct mode.

As pointed out above, a strong indication of the nonclassical nature of the correlations between fields 1,2 is given by $G_{12} = P_c/P_2 > 2$. For this reason, all figures plotting $P_c$ also present the corresponding $P_2$ level, so one may check the condition $G_{12} > 2$ for the measured data. A measurement of the dependence of $g_2^c$ with $G_{12}$ is provided in~\cite{Laurat2006} for experimental conditions close to ours. There, it is directly verified that typically $g_2^c < 1$ for $G_{12} > 2$. In Fig.~\ref{fig2}, the dashed blue line connects the measured values of $P_2$, with the small error bars omitted for clarity, providing then the level to which $P_c$ should be compared in order to determine its degree of correlation to field 1. We observe a threshold to quantum correlations for $OD \approx 0.6$. 

A log-log graph for the increment of the signal above its background ($\Delta P_c = P_c - P_2$) is shown as an inset in Fig.~\ref{fig2}. We note then that $\Delta P_c \propto OD^s \propto N^s$ with $s = 1.9 \pm 0.3$ around threshold, consistent with the expected $s=2$ for a typical superradiant signal. This fast growth of $\Delta P_c$, however, decreases as $P_c$ enters the saturation region, where the probability to extract the single stored excitation becomes closer to one (as pointed out above $P_c \approx 10\%$ indicates a probability of $\approx 50\%$ to extract the single photon from the ensemble in the correct mode).

\subsection{Decoherence}

\begin{figure}[htb]
  \hspace{0.0cm}\includegraphics[width=6.0cm]{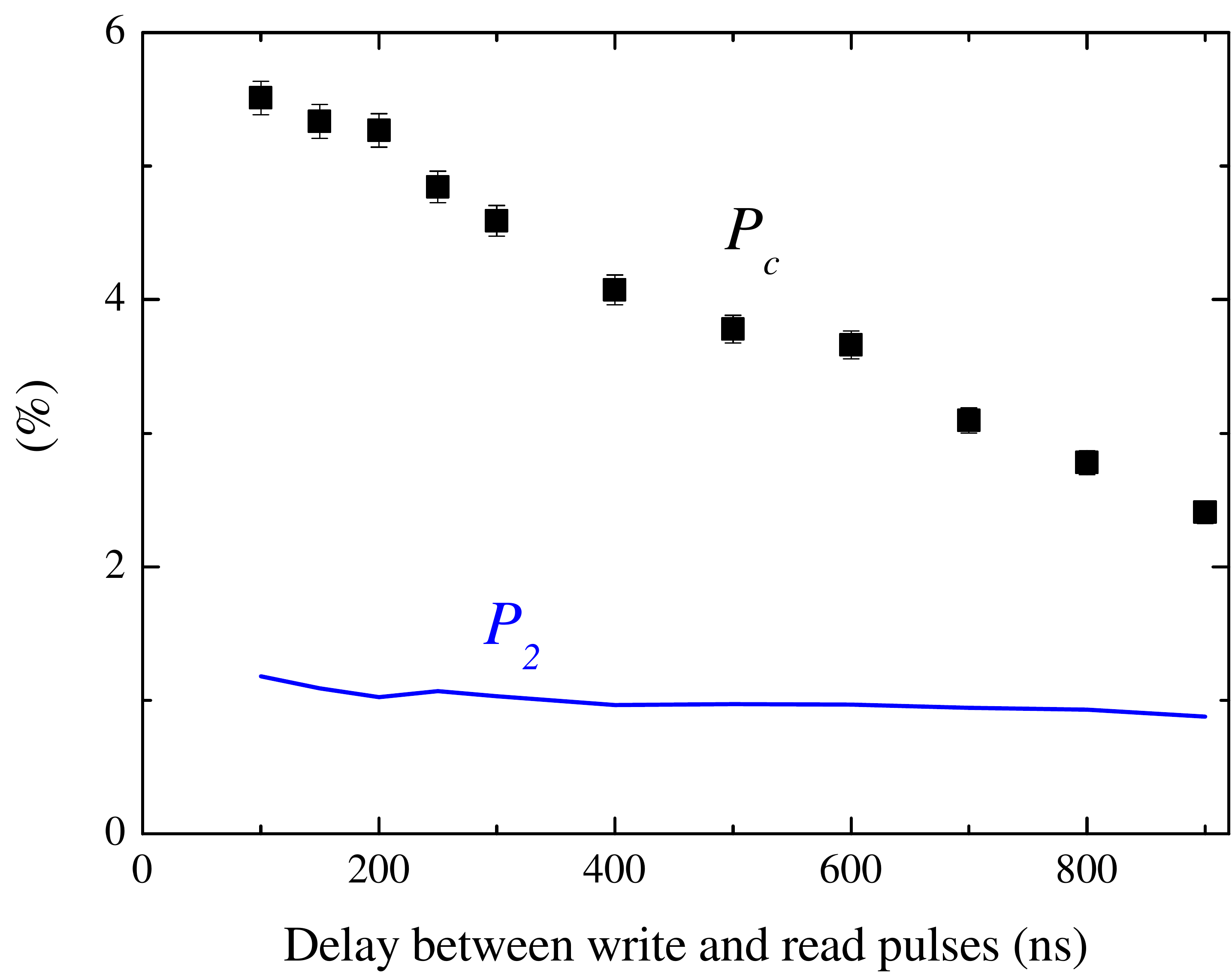}
  \vspace{-0.2cm}
  \caption{Conditional $P_c$ and unconditional $P_2$ probability of detecting the second photon as a function of the delay between write and read pulses.} \label{figS3}
\end{figure}

The quantities introduced above to characterize the correlation between fields 1,2 may also be used to measure the coherence time between levels $|g\rangle$ and $|s\rangle$.Three pairs of bias coils in Helmholtz configuration are employed to cancel spurious DC magnetic fields after the trap magnetic field is turned OFF. After this cancelation, a coherence time of more than 500~ns was observed through the measurement of $P_c$ (squares) as a function of the delay between write and read pulses (Fig.~\ref{figS3}). This coherence time is much larger than the duration of the wavepacket of the second photon (as measured in the following sections), allowing us to neglect decoherence effects when comparing the experimental results to the theory. In Fig.~\ref{figS3}, the blue line plots $P_2$, so that we observe that the strong correlations between fields 1 and 2 also live for more than 500~ns.
  
\vspace*{-0.2cm}

\section{Collective Rabi oscillations}
\label{IV}

The wavepackets of photon 2 are obtained from the quantity $p_c(t) = p_{12}(t)/P_1$, with $p_{12}(t)$ the joint probability of detecting an event in field 1 and another in field 2 in a time window $\Delta t$ around $t$, with $t=0$ the moment the read field is turned on. In this way, $p_c(t)$ provides the conditional probability of detecting an event in field 2 around $t$, after an event in field 1 heralding the preparation of the proper collective state. From this definition, we have then $P_c = \int_0^{\infty}p_c(t)dt$. The unconditional wavepackets of photon 2 are obtained from $p_2(t)$, the probability to detect an event in field 2 in a time window $\Delta t$ around $t$. The time dependence of the correlations between fields 1,2 is given by $g_{12}(t) = p_c(t)/p_2(t) = p_{12}(t)/[p_2(t)P_1]$, with $g_{12}(t) > 2$ again indicating nonclassical correlations~\cite{Polyakov2004}.

\begin{figure}[htb]
  \hspace{0.0cm}\includegraphics[width=9.0cm]{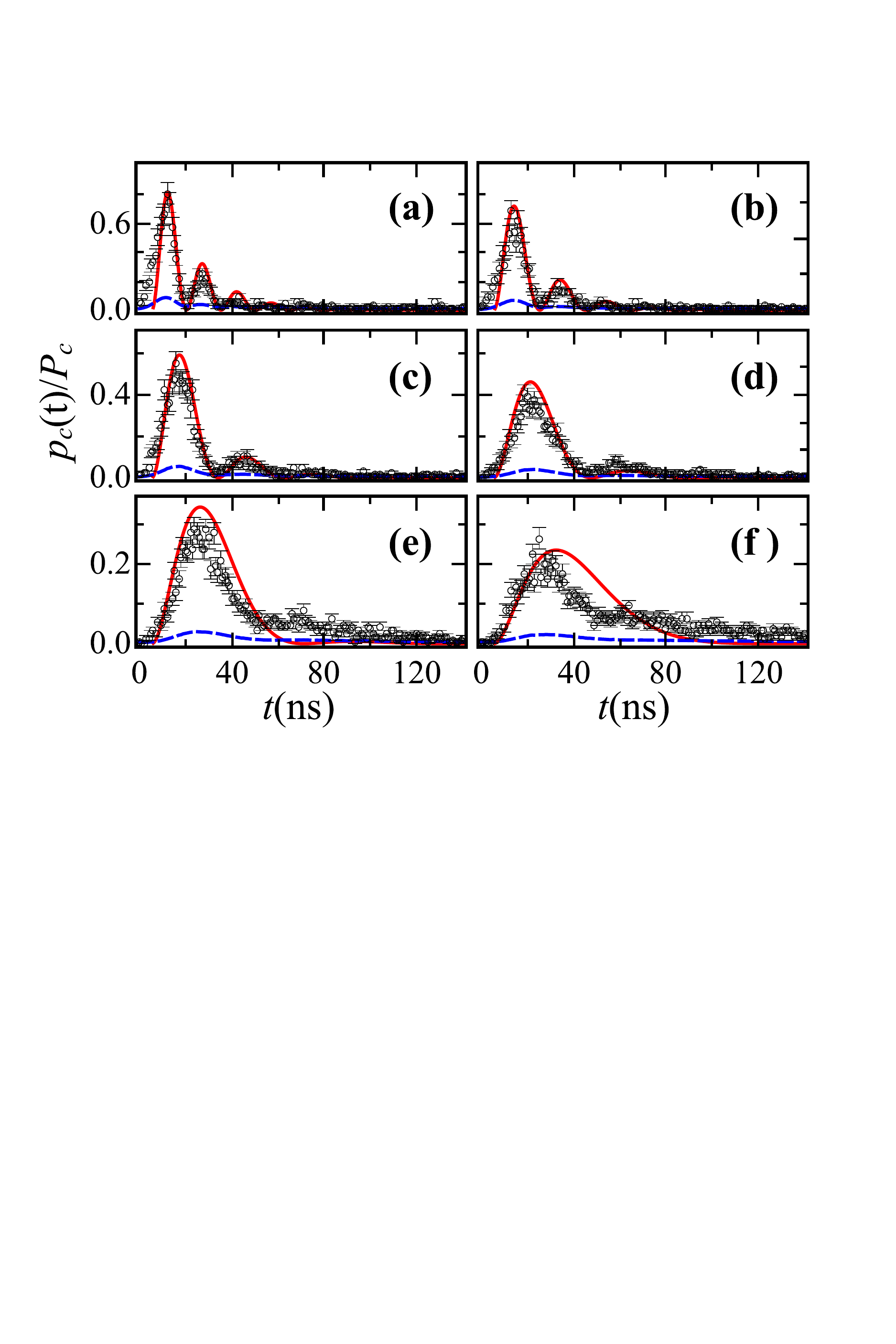}
  \vspace{-6.8cm}
  \caption{Open circles provide the normalized conditional probability as a function of time for six different read powers (in mW): (a) 2.1, (b) 1.2, (c) 0.6, (d) 0.3, (e) 0.15, (f) 0.075. Solid red curves are the theoretical results of Eq.~(\ref{pcPc}) with $\chi = 3.8$ and $\alpha = 9.0$. Dashed blue lines provide the corresponding results for the normalized unconditional probability $p_2(t)/P_c$.}\label{fig3}
\end{figure}

The wavepackets of photon 2 for $OD \approx 4.8$, $\Delta t = 1$~ns and various read powers are plotted in Fig.~\ref{fig3} (open circles). In order to decrease the number of free parameters in comparisons to the theory, we normalized $p_c(t)$ by $P_c$ for each curve. The time dependence for the correlations between fields 1,2 can be evaluated (similarly as for Fig.~\ref{fig2}) by the ratio of the open circles to the dashed blue curves in Fig.~\ref{fig3}~\cite{Polyakov2004}. For these results on the field-2 wavepackets, we introduced some modifications on the timing scheme of Fig.~\ref{figS1}. The period in which the trials were taken was reduced from 1~ms to 0.5~ms, to improve the uniformity of $OD$ throughout the magnetic-field-off period (see subsection below). The trial period was reduced from 5~$\mu$s to 1~$\mu$s to increase the experiment's repetition rate. Finally, the read pulse duration was increased from 180~ns to 840~ns, to guarantee the depletion of the $|s\rangle$ level even for the lowest read powers employed in the measurements.

The theoretical expression for the wavepacket (solid red curves) can be directly obtained from~\cite{Mendes2013} for a resonant read field in the limit of high intensity and negligible decoherence rate between the ground states:
\vspace{-0.1cm}
\begin{equation}
\frac{p_c(t)}{P_c} = \frac{\chi\Gamma \Omega^2 \Delta t \, e^{-\chi\Gamma t/2} }{\left( \Omega^2 - \frac{\chi^2\Gamma^2}{4} \right)} \sin^2\left( \sqrt{\Omega^2 - \frac{\chi^2\Gamma^2}{4}}\, \frac{t}{2}\right) , \label{pcPc}
\end{equation}

\noindent
with $\Gamma = 5.2$~MHz the natural decay rate of the excited state, $\Omega$ the Rabi frequency for the transition $|s\rangle \rightarrow |e\rangle$ excited by the read laser, and $\chi$ a ``cooperativity parameter'' leading to the enhanced decay rates characteristic of superradiance. In~\cite{Mendes2013}, it was also shown that, for our experimental situation, 
\vspace{-0.2cm}
\begin{equation} 
\chi = 1 + \frac{N}{w_0^2k_2^2} \;, \label{chi}
\end{equation}

\noindent
with $w_0$ the waist of the photonic transversal mode, $k_2$ the modulus of its wavevector, and $N$ the number of atoms that interact with this mode. Since both $|s\rangle$ and $|e\rangle$ have Zeeman sublevels, $\Omega$ represents only an effective Rabi frequency. As $\Omega^2$ is proportional to the intensity of the read beam, one may write $\Omega = \alpha \sqrt{P} \Gamma$, with $P$ the read power and $\alpha$ a fit parameter. In this way, we are left with just two fit parameters for all curves: $\alpha$ and $\chi$.  

\vspace{-1.4cm}

The observed behavior in Fig.~\ref{fig3} for the highest powers can be described then as a damped, forced oscillation, with a straightforward physical interpretation from Eq.~(\ref{pcPc})~\cite{Mendes2013}. The read beam induces transitions between levels $|s\rangle$ and $|e\rangle$, forcing the system to perform Rabi oscillations between these levels. The frequency of these oscillations is determined mainly by the strength of the read field, represented by $\Omega$, but also depends on $\chi\Gamma$. When an atom is in $|e\rangle$, it may spontaneously decay to $|g\rangle$ emitting a photon. The spontaneous emission rate is increased due to constructive interference from the emission of different atoms, with such increase represented by the parameter $\chi$. A direct consequence of the superradiant nature of the emission is then the observation of $\chi >1$. For comparison, the natural decay time of independent atoms for the coherence between excited and ground states is $(\Gamma/2)^{-1} \approx 60$~ns, while Fig.~\ref{fig3} shows a decay time of $(\chi\Gamma/2)^{-1} \approx 16$~ns. We observe a better agreement between theory and experiment for higher powers, since the theory in~\cite{Mendes2013} was deduced assuming transparency of the sample to the outgoing photon, due to EIT. As power is reduced, the medium becomes more opaque to the photon and propagation effects related to its reabsorption cannot be neglected~\cite{Mendes2013,deOliveira2012}.
 
\vspace{-0.0cm}
\subsection{OD uniformity}

\begin{figure}[htb]
  \hspace{0.0cm}\includegraphics[width=6.0cm]{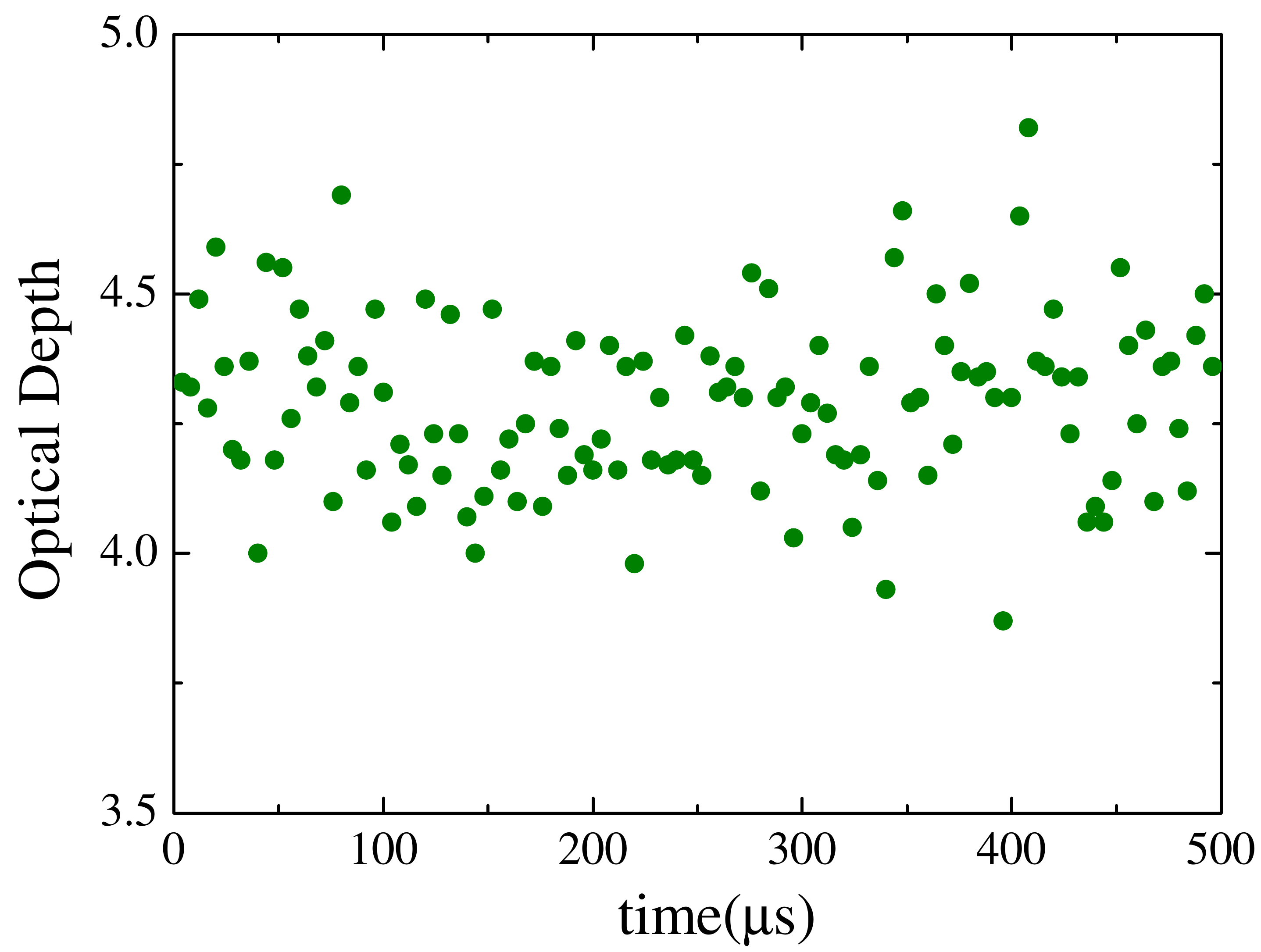}
  \vspace{-0.2cm}
  \caption{Measurement of optical depth for the outgoing photon throughout the whole 0.5-ms interval at which trials are taken with the trap off.} \label{figS2}
\end{figure}

In order to check for the uniformity of $OD$ throughout the whole interval the magnetic field is off, we developed a different method to measure $OD$ in a situation the closest possible to our actual experiment. It consisted in tuning the write field to resonance, considerably decreasing its power, and checking at each trial for the pulse shape after the cell with and without atomic cloud. The distorted pulse shape was then compared to the theoretical result for the propagation of an optical pulse of similar duration through an ensemble of two-level atoms with the same linewidth of the excited state in our experiment. $OD$ was then obtained from the fit of the theory to the experimental results. An example of such measurement is shown in Fig.~\ref{figS2}, for the timing employed in the measurements of Fig.~\ref{fig3}. Figure~\ref{figS2} shows then the measurement of $OD$ throughout the complete 0.5~ms interval in which the trials occur each time the magneto-optical trap is turned off. Each point in the graph is an average for the fit performed in four trials. In this way, we clearly observe no trend for change in the average value of $OD$ during the period the trap is off and the measurements are performed. The average $OD$ over the whole 0.5-ms interval measured this way, $OD = 4.29\pm 0.01$, was also consistent with the $OD$ obtained from our standard method under the same conditions. 

\section{Variation of decay time with optical depth}
\label{V}

\begin{figure}[htb]
  \hspace{0.5cm}\includegraphics[width=9.0cm]{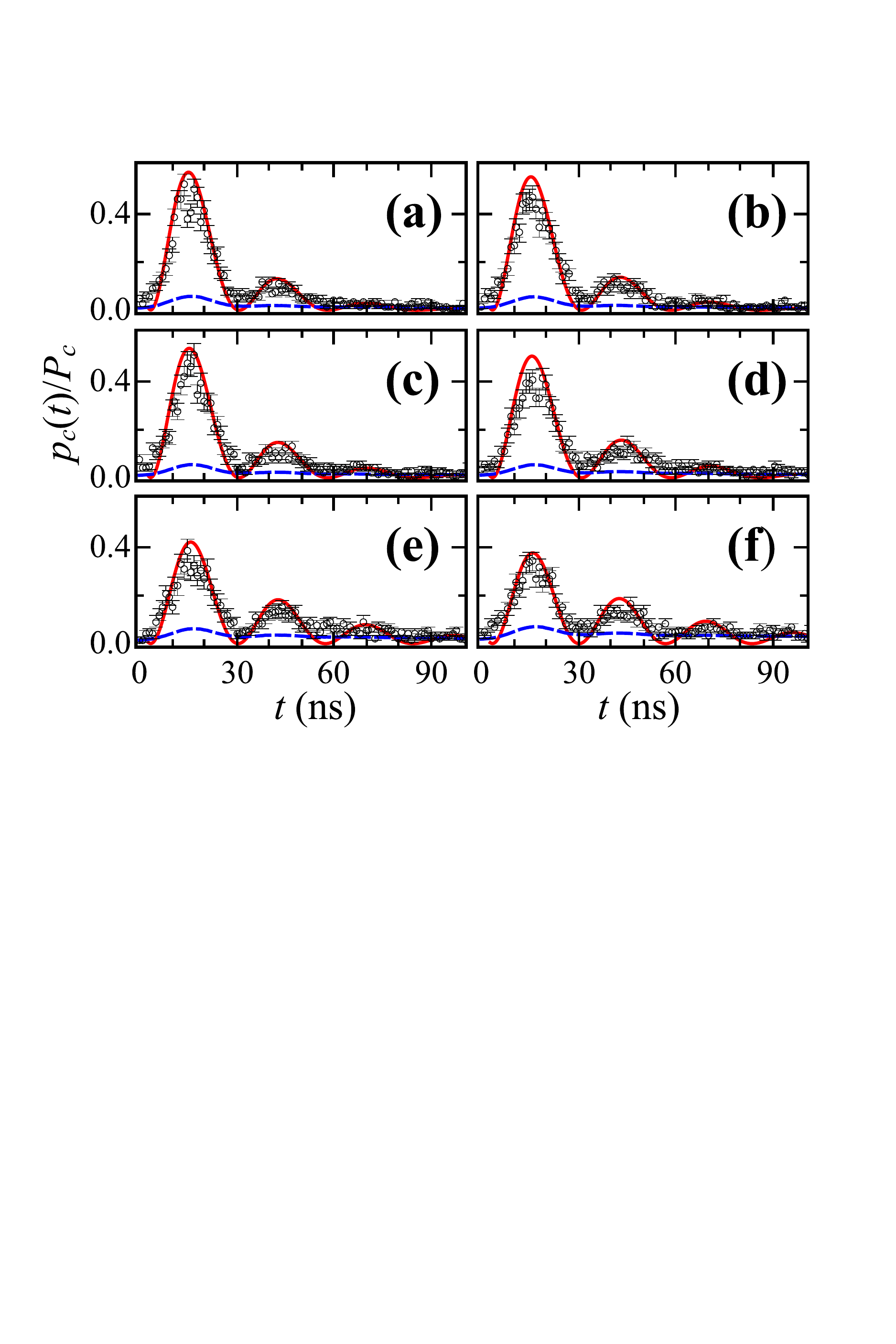}
  \vspace{-6.7cm}
  \caption{Open circles provide the normalized conditional probability as a function of time for six different Optical Depths: (a) 4.8, (b) 4.0, (c) 3.4, (d) 2.6, (e) 1.6, (f) 1.0. Solid red curves are the theoretical results for independent fits using Eq.~(\ref{pcPc}). The fit parameters $\chi$ and $\alpha$ for each OD are plotted in Fig.~\ref{fig5}. Dashed blue lines provide the corresponding results for the normalized unconditional probability $p_2(t)/P_c$. The read power is 0.3~mW.}\label{fig4}
\end{figure}

\begin{figure}[htb]
  \hspace{0.0cm}\includegraphics[width=7.5cm]{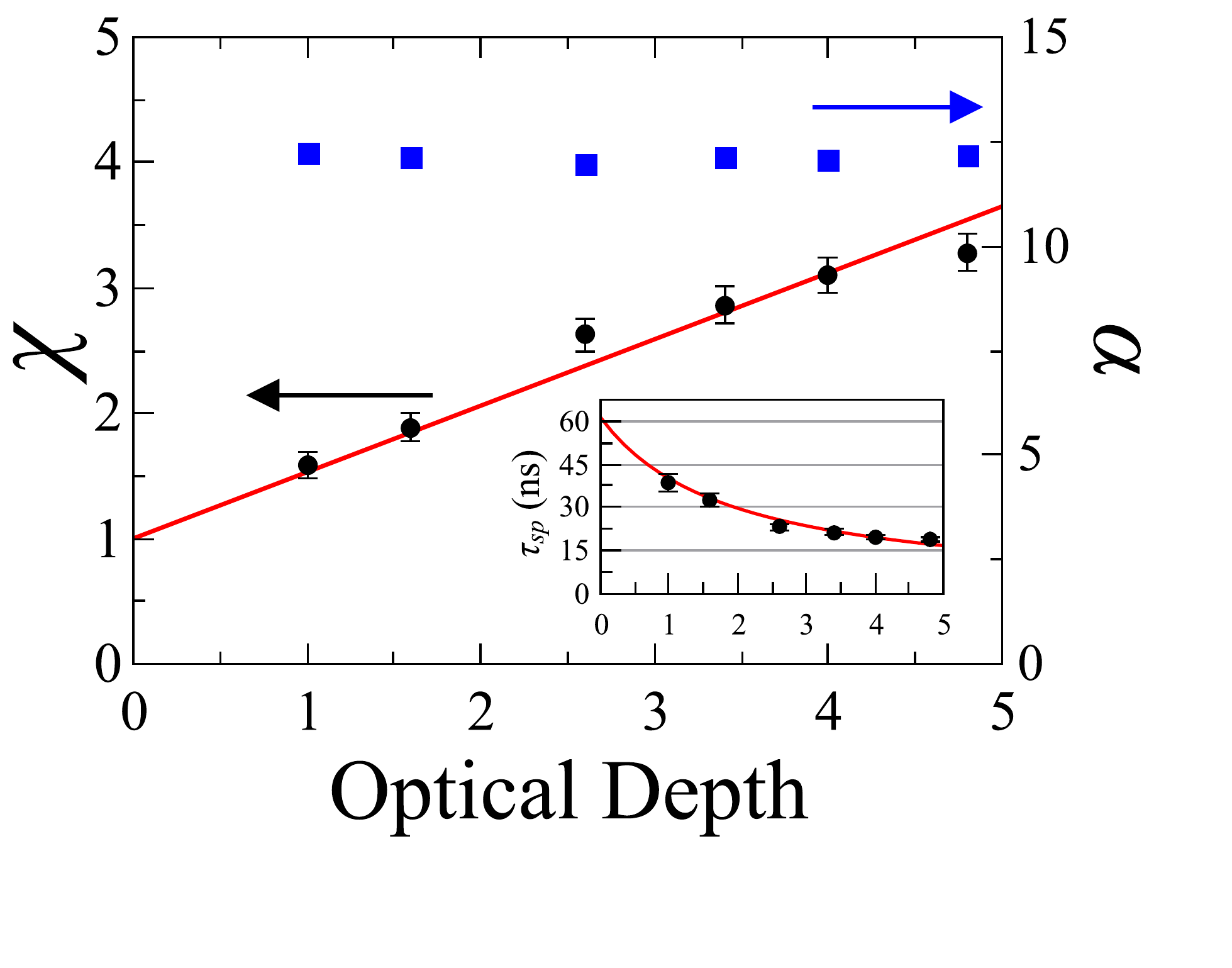}
  \vspace{-1.0cm}
  \caption{Values of $\chi$ and $\alpha$ as a function of $OD$ obtained from the fittings in Fig.~\ref{fig4}. The red curve is a fit to a function with the same linear dependence as Eq.~(\ref{chi}). The inset plot the decay times $\tau_{sp} = (\chi \Gamma/2)^{-1}$.}\label{fig5}
\end{figure}

Differently from the case of independent atoms, the decay rate in superradiance may vary by tuning the number of atoms in the ensemble, as expected from Eq.~(\ref{chi}). Our results for the wavepacket of photon 2 as $OD$ is tuned are plotted in Fig.~\ref{fig4}. The symbols and colors in this figure are the same as for Fig.~\ref{fig3}. The read power is around $0.3$~mW. The theoretical plots were obtained from independent fittings of the experimental data to Eq.~(\ref{pcPc}), using $\chi$ and $\alpha$ as fitting parameters. The values of $\chi$ and $\alpha$ obtained for each $OD$ are plotted in Fig.~\ref{fig5}. Since $OD$ is proportional to $N$, following Eq.~(\ref{chi}) we finally fit the data for $\chi$ versus $OD$ with the curve $\chi = 1 + \beta \, OD$, obtaining $\beta = 0.53 \pm 0.02$~(see subsection below). We can see that the increase in the spontaneous decay rate of the atomic ensemble is proportional to its number of atoms, characterizing the superradiant nature of the observed single-photon emission. In Fig.~\ref{fig4} we were able to change the decay time from a minimum of $18.6\pm 0.8$~ns in panel (a) to about $38\pm 3$~ns in (f) (see inset in Fig.~\ref{fig5}). As the number of atoms becomes too low, however, the visibility of the Rabi oscillations degrades due to the increase in the noise floor given by $p_2(t)$ (dashed blue lines). This is expected from the behavior of $P_c$ as $OD$ decreases (Fig.~\ref{fig2}), since it eventually reaches the noise floor given by $P_2$ once the collective enhancement is lost. 
  
\vspace*{-0.0cm}
  
\vspace{-0.5cm}
\subsection{$\beta$ parameter}  

The optical depth can be written as $OD = \alpha_0 l$, with $\alpha_0$ the optical density of the sample and $l$ its length. On the other hand, we have $\alpha_0 = \sigma_0 N/V$, where $\sigma_0$ is the on-resonance scattering cross section of the atom and $V$ is the sample's volume. In this way, we have $OD = N \sigma_0 / \pi w_0^2$, with $w_0$ the waist of the photonic gaussian mode defining the ensemble's volume. The interpretation for this expression is straightforward, since each atom will scatter a portion $\sigma_0 / \pi w_0^2$ of the incident beam resulting in the total thickness of the sample when the contributions of all atoms are combined.

The value $\beta = 0.53\pm 0.02$ was obtained from a fit of the function $\chi = 1 + \beta OD$ to the experimental data. From the above relation between $OD$ and $N$ and from Eq. (2), we obtain then
\begin{equation}
\beta = \frac{\pi}{\sigma_0k^2} \;,
\end{equation}
with $k$ the wave vector of field 2. The value of $\sigma_0$ can be obtained from $\sigma_0 = \hbar \omega \Gamma / 2 I_{sat}$, with $\omega$ the optical\linebreak 

\vspace*{-0.0cm}
    
\vspace{-0.2cm}

\noindent frequency of the transition and $I_{sat}$ its saturation intensity~\cite{Steck}. From this simple analysis, $\beta$ would be given by various well known parameters ($k$,$\omega$,$w_0$,$\Gamma$) plus the saturation intensity $I_{sat}$, which will be given as an average over various transitions through different Zeeman sublevels. However, the observed value of $\beta$ would imply a saturation intensity of $I_{sat} \approx 3.5$~mW/cm$^2$, a reasonable value for this specific transition of cesium~\cite{Steck}.   

\vspace{0.3cm}
\section{Conclusions}
\label{VI}

We presented an investigation on the superradiant nature of the DLCZ photon-pair-generation process, demonstrating the observation of single-photon superradiance in the system. Various aspects of superradiance were then demonstrated in connection with the single-photon emission. Particularly, we measured the threshold optical depth of the ensemble to start the superradiant process, and demonstrate that the probability to emit the single photon grows with the square of the number of atoms at threshold. We also measure the decrease of the decay time from the excited state, for single-photon emission, as the number of atoms is increased, following the expected behavior for superradiance. Key to our approach is a close comparison of our experimental data for the photonics wavepacket to an analytical theory for the reading process obtained from first principles~\cite{Mendes2013}.  

Finally, the presented investigation opens the way for a systematic, bottom-up approach to the study of superradiance. The minimal superradiant cascade implemented up to now may be expanded using quantum-optics techniques of conditional generation of states with a larger number of excitations~\cite{Ourjoumtsev2006}. In this way, it should be possible to generate superradiant cascades with two or more excitations, controllably moving down the typical de-excitation ladder of macroscopic superradiance.  

\begin{center}
{\bf ACKNOWLEDGEMENTS} 
\end{center}

\vspace{-0.2cm}
We gratefully acknowledge K. N. Cassemiro for her assistance in part of the experiment. This work was supported by CNPq, CAPES, PRPq/UFMG, and FACEPE (Brazilian agencies), particularly through the programs PRONEX and INCT-IQ (Instituto Nacional de Ci\^encia e Tecnologia
de Informa{\c c}\~ao Qu\^antica).

\end{document}